\documentstyle[12pt,bezier]{article}
\newcommand{\be}{\begin{equation}}
\newcommand{\ee}{\end{equation}}

\renewcommand{\vec}[1]{{\bf #1}}
\begin{document}
\begin{flushright}
SUNY-NTG-95-34
\end{flushright}
\begin{center}
{\Large Thermal phase transition in QCD}
\footnote{Lecture on the International School of Physics ``Enrico Fermi ''
(Varenna, July 1995).}
\vspace{1cm}

A.V. Smilga\\
\vspace{0.5cm}

{\it Department of Physics, SUNY at Stony Brook, Stony Brook, NY 11794-3800,
USA}
\footnote
{Permanent address: ITEP, B. Cheremushkinskaya 25, Moscow 117259, Russia.}
\vspace{1cm}
\end{center}

\abstract{We give a review of modern theoretical
understanding of the problem
of  thermal phase transition in $QCD$. The existence or
non-existence of such transition
depends on the nature of the gauge group, the number of
light quark flavors
, and on the value of quark masses. Numerical lattice
measurements indicate
that the phase transition {\it does} not occur for the
experimentally observed
values of quark masses.}

\section{Introduction.}
The properties of $QCD$ medium at finite temperature have been
the subject of intense study during the last 15 years. It was
realized that the properties of the medium undergo a drastic
change as the temperature increases. At low temperatures, the
system presents a gas of colorless hadron states --- the
eigenstates of the $QCD$ hamiltonian at zero temperature. When
the temperature is small, this gas is composed mainly of pions
--- other mesons and baryons have higher mass and their
admixture in the medium is exponentially small $\sim
\exp\{-M/T\}$. At small temperature, also the pion density is
small --- the gas is rarefied and pions practically do not
interact with each other.

 However, when the temperature increases, pion density grows,
the interaction becomes strong, and also other strongly
interacting hadrons appear in the medium. For temperatures of
order $T \sim$ 150 Mev and higher, the interaction becomes so
strong that the hadron states do not present a convenient basis
to describes the properties of the medium anymore, and no
analytic calculation is possible.

 On the other hand, when the temperature is very high, much
higher than the characteristic hadron scale $\mu_{hadr} \sim$
0.5 Gev, theoretical analysis becomes possible again. Only in
this range, the proper basis are not hadron states but quarks
and gluons --- the elementary fields entering the $QCD$
lagrangian. For high temperatures, a characteristic energy of
quarks and gluons travelling through the medium is also high,
the effective coupling constant is small, and the system
presents the {\it quark-gluon plasma}.  Its properties are in
many respects very similar to the properties of the usual
non-relativistic plasma involving charged particles with weak
Coulomb interaction. The only difference is that quarks and
gluons carry not the electric, but color charge. The
perturbation theory in the coupling constant can be developed
and many thermodynamic (such as free energy) and kinetic (such
as viscosity) characteristics of the medium can be analytically
evaluated \cite{QGP}.

 Thus, the properties of the system at low and at high
temperatures have nothing in common. A natural question arises:
What is the nature of the transition from low-temperature hadron
gas to high- temperature quark-gluon plasma ? Is it a {\it
phase} transition ? If yes, what is its order ?  I want to
emphasize that this question is highly non-trivial. A drastic
change in the properties of the system in a certain temperature
range does not guarantee the presence of the phase transition
{\it point} where free energy of the system or its specific heat
is discontinuous. Recall that there is no phase transition
between  ordinary gas and ordinary plasma.

 We shall see in the following that, as far as the real $QCD$
with particular values of quark masses is concerned, the answer
is probably negative. What really happens is not the phase
transition but a sharp crossover --- "almost" a second-order
phase transition.  However, the real phase transition {\it does}
occur in some relative theories --- in pure Yang- Mills theory
(when the quark masses are sent to infinity) and in $QCD$ with 2
or 3 exactly massless quark flavors.

 There are at least 4 reasons why this question is interesting
to study: \begin{enumerate} \item It is just an amusing
theoretical question.  \item Theoretic conclusions can be
checked in lattice numerical experiments.  Scores of papers
devoted to lattice study of thermal properties of QCD have been
published.  \item Perhaps, a direct experimental study would be
possible on RHIC --- high-energy ion collider which is now under
construction.  I'll discuss the possibility to observe a
beautiful effect, the so called {\it disoriented chiral
condensate} in the end of the lecture.  \item During the first
second of its evolution, our Universe passed through the stage
of high-$T$ quark-gluon plasma which later cooled down to hadron
gas (and eventually to dust and stars, of course). It is
essential to understand whether the phase transition did occur
at that time. A profound first-order phase transition would lead
to observable effects. We know (or almost know --- the
discussion of this question has not yet completely died away)
that there were no such transition. But it is important to
understand why.
\end{enumerate}
 Note that there is also a related but {\it different} question
--- what are the properties of relatively cold but very dense
matter and whether there is a phase transition when the chemical
potential corresponding to the baryon charge rather than the
temperature is increased.  This lecture will be devoted
exclusively to the thermal properties of QCD, and we shall
assume zero baryon charge density.

 \section{Pure Yang-Mills theory: deconfinement phase
transition.}
\setcounter{equation}0
 This is the system where the phase transition from the glueball
phase to the gluon plasma phase {\it does} occur. This result
has been obtained long ago by Polyakov \cite{Pol} and Susskind
\cite{Sus}. On the heuristic level, the reasoning is the
following:

 We know (for real $QCD$ --- from experiment, and for pure YM
theory --- from theoretical arguments and from lattice
measurements) that the theory  enjoys confinement at low
temperature. That means that the potential between the test
heavy quark and antiquark grows linearly at large distances:
\be
\label{conf}
T=0:\ \ \ V_{Q\bar{Q}}(r) \sim \sigma r,\ \ \ r
\rightarrow
\infty
 \ee
 On the other hand, at high temperature when the system presents
a weakly interacting plasma of gluons, the behavior of the
potential is quite different:

\be \label{Deb}
T \gg
\mu_{hadr}: \ \ \ V_{Q\bar{Q}}(r) \sim \frac {g^2(T)}{r} e^{-m_D
r}
\ee
Here $m_D \sim gT$ is the Debye mass, and the potential is the
Debye screened potential much similar to the usual Debye
potential between static quarks in non-relativistic plasma.
There is no confinement at large $T$. There should be some point
$T_c$ (the critical temperature) where the large $r$ asymptotics
of the potential changes and the phase transition from the
confinement phase to the Debye screening phase occurs.

   These simple arguments can be formulated in a rigorous way.
Consider the partition function of the system written as the
Euclidean path integral. It is known since Matsubara that, at
finite $T$, the fields are defined on the cylinder: Euclidean
time $\tau$ lies within the range $0 \leq \tau \leq \beta = 1/T$
, and one should impose periodic boundary conditions on the
gluon fields:
\be \label{bc}
A_\mu^a(\vec{x}, \beta) =
A_\mu^a(\vec{x}, 0)
\ee
Let us choose a gauge where $A_0^a$ is
time-independent.  Introduce the quantity called the Polyakov
loop
\be \label{P}
P(\vec{x}) = \frac 1{N_c} {\rm Tr} \exp\{
ig\beta A_0^a(\vec{x}) t^a\}
\ee
It is just a Wilson loop on the
contour which winds around the cylinder.  Consider the
correlator
\be \label{PP}
C_T(\vec{x}) = <P(\vec{x}) P^*(0)>_T
\ee
One can show \cite{Nad} that the correlator (\ref{PP}) is
related to the free energy of the test heavy quark-antiquark
pair immersed in the plasma.
\be \label{FQQ}
 C_T(\vec{x}) =
\frac 34 \exp\{-\beta F_{Q\bar{Q}}^{(3)}(r)\} + \frac 14 \exp \{
-\beta F_{Q\bar{Q}}^{(0)}(r)\}
\ee
where $r = |\vec{x}|$.
$F_{Q\bar{Q}}^{(3)}(r)$ and $F_{Q\bar{Q}}^{(0)}(r)$ are free
energies of test quark-antiquark pairs (alias static potentials)
in the triplet and, correspondingly, the singlet net color
state. Let us take now the limit $r \rightarrow \infty$. The
quantity
\be \label{Cinf}
C_T(\infty) = \lim_{r \rightarrow
\infty} C_T(r)
\ee
plays the role of the {\it order parameter}
of the deconfinement phase transition. At small $T$,
$F_{Q\bar{Q}}(r)$  grow linearly at  $r \rightarrow
\infty$ and $C_T(\infty) = 0$. At large $T$, free energies
do not grow and $C_T(\infty)$ is some non-zero constant (if one
would naively substitute in Eq.(\ref{FQQ}) the Debye form of the
potentials (\ref{Deb}), one would get $C_{T \gg
\mu_{hadr}}(\infty) = 1$, but it is not quite true because
$F_{Q\bar{Q}}(r)$ involve also a constant depending on the
ultraviolet cutoff of the theory. See \cite{Jengo,bub} for
detailed discussion). There is a phase transition in between.

What are the properties of this phase transition ? There are not
quite rigorous but suggestive theoretical arguments based on the
notion of ``universality class'' \cite{Svet} which predict
different properties for different gauge groups. The main idea
is that the pure YM theory based on $SU(2)$ color group has some
common features with the Ising model (with global symmetry
$Z_2$), the theory with $SU(3)$ gauge group --- with a
generalized Ising model (the Potts model) with the global
symmetry $Z_3$ etc.  The Ising model has the second order phase
transition, and the same should be true for pure $SU(2)$ gauge
theory. Systems with $Z_N$ symmetry display, however, the first
order phase transition, and the same should be true for pure
$SU(N \geq 3)$ theory. The lattice data
\cite{purlat}
are in a nice agreement with this prediction. Also, critical
indices of the second order phase transition were measured.
Their numerical values are close to the numerical values of
critical indices in the Ising model.

\subsection{Bubble confusion.}
There was a long-standing confusion concerning the nature of
deconfinement phase transition in pure YM theory. It has been
clarified only recently and I want to dwell on this question in
more details.

In scores of papers published since 1978, it was explicitly or
implicitly assumed that one can use the cluster decomposition
for the correlator (\ref{PP}) at large $T$ and attribute the
meaning to the temperature average $<P>_T$. Under this
assumption, the phase of this average can acquire $N_c$
different values: $<P>_T = C\exp\{2\pi ik/N_c\}, \ \ k =
0,1,\ldots, N_c-1$ which would correspond to $N_c$ distinct
physical phases and to the spontaneous breaking of the discrete
$Z_N$-symmetry. In recent \cite{Pisa}, the surface energy
density of the domain walls separating these phases has been
evaluated.

However, the standard interpretation is wrong. In particular:
\begin{enumerate}
\item Only the correlator (\ref{PP}) has the physical
meaning. The phase of the expectation value $<P>_T$ is not a
physically measurable quantity. There is only {\it one} physical
phase in the hot YM system.
\item The ``walls'' found in \cite{Pisa} should not be
interpreted as physical objects living in Minkowski space, but
rather as Euclidean field configurations, kind of ``planar
instantons'' appearing due to non-trivial $\pi_1(\cal G) = Z_N$
where ${\cal G} = SU(N)/Z_N$ is the true gauge symmetry group of
the {\it pure} Yang-Mills system.
\item The whole bunch of arguments which is usually applied
to non-abelian gauge theories can be transferred with a little
change to hot $QED$. The latter also involves planar instantons
appearing due to non- trivial $\pi_1[U(1)] = Z$. These
instantons should {\it not} be interpreted as Minkowski space
walls.
\end{enumerate}

It is impossible to present an adequate discussion of this issue
in this short lecture. We refer the reader to \cite{bub} where
such discussion was given. Here we restrict ourselves by
outlining some heuristic physical arguments.

Right from the beginning, one meets a puzzle. In standard
approach, $Z_N$ is broken spontaneously at {\it high}
temperature and restored at {\it low} temperature. This is very
strange and unusual. The opposite is much more common in
physics. There are some models where symmetry breaking survives
and can even be induced at high temperature \cite{Moha}, but the
mechanism of this breaking is quite different from what can
possibly occur in the pure YM theory.

The second observation is that $Z_N$ symmetry which is
presumably broken in the deconfinement phase is just not there
in the continuum theory: the gluon fields are not transformed
under the action of the center. $Z_N$ - symmetry is present in
the standard lattice version of the theory but is absent, again,
in the lattice theory involving adjoint matrices $O^{ab} = {\rm
Tr} \{t^a U t^b U^\dagger\}$ rather than the unitary matrices
$U$.

It was mentioned earlier that the Polyakov loop expectation
value $<P>_T$ as such has no physical meaning. Let us explain
why. Being taken at face value, $<P>_T$ would measure the free
energy of a single fundamental heavy source immersed in the
system \cite{Larry}: $<P>_T = \exp\{-\beta F_T\}$. As a matter
of fact, nonzero phase of $<P>_T$ would correspond to the
complex free energy which is an obvious nonsense. But the point
is that one just cannot put a single fundamental source into the
system due to the Gauss law constraint \cite{Hift}. The net
color charge should be zero, and a fundamental source cannot be
screened by gluons --- the only dynamic fields in the lagrangian
of the theory and in the heat bath
\footnote{To be quite precise, the net color charge can be made
non-zero due to a boundary term at spatial infinity. In a finite
spatial box, such a term can appear when non-standard
(non-periodic) boundary conditions are chosen. But then the
phase of $<P>_T$ would be exactly determined by these boundary
conditions. As the physical properties of the theory cannot
depend on b.c.  when the box is large enough, it is just another
way to say that the phase is not physical.}

What one can well do is to immerse a heavy quark-antiquark pair
and measure thereby the correlator (\ref{PP}) which {\it is}
physical.  Or, say, for the $SU(3)$ gauge group, one can immerse
3 heavy quarks at different points and measure $<P(\vec{x})
P(\vec{y}) P(0)>_T$ which is physical, again, but, in contrast
to $<P>_T$, does not involve the phase uncertainty.

Actually, the delusion of spontaneous $Z_N$ breaking in pure YM
theory persisted for so long because people habitually described
this system in terms of $A^a_0(\vec{x})$ (more exactly --- in
terms of $\Omega_{\vec{x}} = \exp\{i\beta g
A_0^a(\vec{x})t^a\}$) which are not the dynamic variables
entering the hamiltonian but the variables dual to the Gauss law
constraints.  A close analogy can be drawn with the Ising model
in 2 dimensions. When expressed in terms of the original spin
variables $\sigma_i$, the system is ordered at low temperatures
and disordered at high temperatures --- the spontaneously broken
$Z_2$ - symmetry is restored there. One can make, however, the
Kramers-Wannier transformation and describe the system in terms
of the dual or "disorder" variables $\eta_i$
\cite{KW}. When the
normal temperature is high, the dual temperature is low, and
the dual hamiltonian $H^*[\eta_i]$ describes, indeed, the system
where $Z_2$ symmetry is broken at high temperatures and restored
at low temperatures. But the variables $\eta_i$ are not
physical observables and, as far as any Gedanken physical
experiment is concerned, $Z_2$ - symmetry in the Ising model
{\it is} restored at high temperatures.

To summarize, there is only one physical phase at high $T$.  Its
properties are relatively simple --- it is the weakly
interacting plasma of gluons.  The description in terms of dual
variables is useful for some purposes (e.g.  the universality
class arguments of Ref. \cite{Svet} which predict the order of
the phase transition are based on the dual description), but one
should be very careful not to read out in it something which is
not in Nature.

\section{$QCD$ with massless quarks.}
\setcounter{equation}0
If the theory involves besides gluons also quarks with finite
mass, the static interquark potential $V_{Q\bar{Q}}(r)$ does not
grow at large distances anymore even at $T=0$. Dynamic quarks
screen the potential of static sources. One can visualize this
screening thinking of the color gluon tube stretched between two
static fundamental sources being torn apart in the middle with
the formation of an extra quark-antiquark pair. Thus, in QCD
with quarks, the Wilson loop average has the perimeter rather
than the area law
\footnote{That does not mean that there is no confinement ---
as earlier, only the colorless states are present in the
physical spectrum. But the behavior of the Wilson loop is not a
good signature of confinement anymore.}.  The correlator of two
Polyakov loops (\ref{PP}) tends to a constant at large distances
universally at low and at high temperature, and this correlator
cannot play the role of the order parameter of phase transition.

Still, the phase transition can occur and does occur in some
versions of the theory. It is associated, however, not with
change in behavior of the correlator (\ref{PP}), but with {\it
restoration of chiral symmetry} which is spontaneously broken at
zero temperature.

Consider YM theory with $SU(3)$ color group and involving $N_f$
massless Dirac fermions in the fundamental representation of the
group. The fermion part of the lagrangian is
\be \label{Lf}
L_f
= i \sum_f \bar q_f \gamma_\mu {\cal D}_\mu q_f
\ee
where ${\cal
D}_\mu = \partial_\mu - igA_\mu^a t^a$ is the covariant
derivative. The lagrangian (\ref{Lf}) is invariant under chiral
transformations of fermion fields:
\be \label{chi}
q_{{\small
L,R}} \rightarrow A_{{\small L,R}}\ q_{{\small L,R}}
\ee
where
$q_{{\small L,R}} = \frac 12 (1 \pm \gamma^5)q$ is the flavor
vector with $N_f$ components and $A_{{\small L,R}}$ are two
different $U(N_f)$ matrices. Thus, the symmetry of the classical
lagrangian is $U_L(N_f) \otimes U_R(N_f)$. Not all N\"other
currents corresponding to this symmetry are conserved in the
full quantum theory. It is well known that the divergence of the
singlet axial current $ j^5_\mu = \sum_f \bar q_f
\gamma_\mu \gamma^5 q_f$ is non-zero due to anomaly:
  \be \label{anom}
\partial_\mu j_\mu^5  \ \sim \ g^2
\epsilon^{\mu\nu\alpha\beta} G^a_{\mu\nu} G^a_{\alpha\beta}
 \ee
Thus, the symmetry of quantum theory is $SU_L(N_f) \otimes
SU_R(N_f) \otimes U_V(1)$. It is the experimental fact that (for
$N_f = 2,3$, at least) this symmetry is broken spontaneously
down to $U_V(N_f)$. The order parameter of this breaking is the
chiral quark condensate $<\sum_f \bar q_f q_f>_0$. This spontaneous
breaking leads to appearance of the octet of pseudoscalar
Goldstone   states in the spectrum. Of course, in the real $QCD$
the quarks are not exactly massless, the mass term is not
invariant with respect to the symmetry (\ref{chi}) but only
under $U_V(N_f)$. As a result, in real World we have the octet
of light (but not massless) pseudo- Goldstone pseudoscalar
states ($\pi, K, \eta$). But the small mass of pseudogoldstones
and the large splitting between the massive states of opposite
parity ($\rho/A_1$, etc.)  indicate beyond reasonable doubts
that the exact chiral symmetry (\ref{chi}) would be broken
spontaneously in the massless case. As the masses of the strange
and, especially, of $u$- and $d$- quarks are small \cite{GLmass}
, the mass term in the lagrangian can be treated as
perturbation. E.g. the pion mass satisfies the relation
\be
\label{pimass}
F_\pi^2 m_\pi^2 = (m_u + m_d) |<\bar u u>_0|
\ee
($F_\pi = 93$  Mev is the pion decay constant) and turns to zero
in the chiral limit $m_{u,d} \rightarrow 0$.

It is noteworthy that the symmetry breaking pattern
\be
\label{br3}
SU_L(N_f) \otimes SU_R(N_f) \rightarrow SU_V(N_f)
\ee
depends crucially on the assumption that the gauge group
involves at least 3 colors. For $SU(2)$ color group where quarks
and antiquarks belong to the same representation (the
fundamental representation of the $SU(2)$ group is pseudoreal:
${\bf 2} \equiv {\bf \bar 2}$), the symmetry group of the
lagrangian (\ref{Lf}) is much higher. It is $U(2N_f)$ and
involves also mixing between quarks and antiquarks. $U_A(1)$ -
part of this symmetry is anomalous and the formation of chiral
condensate breaks spontaneously the remaining $SU(2N_f)$   down
to a simplectic group
\cite{simpl}:
  \be
  \label{br2}
SU(2N_f) \rightarrow Sp(2N_f)
  \ee
As a result, $2N_f^2 - N_f -1$ Goldstone bosons living on the
coset space appear. For $N_f =2$, we have not 3 as usual, but 5
``pions''. This fact is important to understand for people who
would wish to study numerically on lattices the spontaneous
chiral symmetry breaking  with $SU(2)$ gauge group.

As far as the thermal properties of the theory are concerned,
the point is that a spontaneously broken symmetry must be
restored under a sufficient heating. There should be a critical
temperature $T_c$ above which the fermion condensate $<\bar q
q>_T$ is zero. This is the temperature of phase transition and
$<\bar q q>_T$ is the order parameter associated with the
transition.

Note that the phenomenon of spontaneous chiral symmetry breaking
is specific for theories with {\it several} light quark flavors.
In the theory with $N_f = 1$ , the non- anomalous part of the
symmetry of the lagrangian is just $U_V(1)$. It stays intact
after adding the mass term and after taking into account the
formation of the condensate $<\bar q q>$. The condensate is
still formed, but it does not correspond to spontaneous breaking
of any symmetry and need not vanish at high temperature. So, it
does not. At high temperatures when the effective coupling is
small, it can be evaluated semiclassically in the instanton
approach
\cite{GPJ,KY}, and one can show that it falls down as a
power of temperature and never reaches zero.

There is {\it no} phase transition in $QCD$ with only one light
or massless flavor. But it {\it does} occur when the number of
massless flavors $N_f$ is 2 or more.

The melting down of quark condensate can be studied analytically
at low temperature when the medium presents a rarefied weakly
interacting gas of pions with low energies.  Their properties
are described by the effective chiral lagrangian
\be
\label{Lchi}
{\cal L} = \frac 14 F_\pi^2 {\rm Tr} \{\partial_\mu
U \partial_\mu U^\dagger\} + \ldots
\ee
 where $U$ is the
$SU(N_f)$ matrix and the dots stand for higher derivative terms
and the terms involving quark masses. When the characteristic
energy and the quark masses are small, the effects due to these
terms are suppressed and a perturbation theory (the {\it chiral
perturbation theory}
\cite {CPT}) can be developed. In \cite{Ger}, the
temperature dependence of $<\bar q q>_T$ has been determined on
the 3-loop level. In the approximation where only the presence
of pions in the heat bath is taken into account and the effects
due to non-zero $m_u$ and $m_d$ are neglected, the result has a
rather simple form

\be \label{qqT}
<\bar q q>_T = <\bar q q>_0
\left[ 1 - \frac {T^2}{8F_\pi^2} - \frac {T^4}{384F_\pi^4} -
\frac {T^6}{288F_\pi^6} \ln
\frac \Lambda T + \ldots \right]
  \ee
The constant $\Lambda$ depends on the higher-derivative terms in
the effective lagrangian and can be fixed from experiment:
$\Lambda \sim 500 \pm 100$  Mev. The dependence (\ref{qqT})
together with the curves where only only the 1 loop correction
$\propto T^2$ and 2 loop correction $\propto T^4$ are taken into account
(please, do not put attention to the ``technical'' curve marked $a^0_2 = 0$)
 is drawn in Fig. 1 taken from Ref. \cite{Ger}.

\newpage
\vspace{13cm}
The expansion in the parameter $\sim T^2/8F_\pi^2$ makes sense
when this parameter is small, i.e. when $T \leq 100- 150$ Mev.
Strictly speaking, one cannot extrapolate the dependence
(\ref{qqT}) for larger temperatures, especially having in mind
that, at $T > 150$ Mev,  the heat bath includes a considerables
fraction of other than pion hadron states. But as we know anyhow
that the phase transition with restoration of chiral symmetry
should occur, the estimate of the phase transition temperature
(i.e. the temperature when $<\bar q q>_T$ hits zero) based on
such an extrapolation is not altogether stupid. This estimate is

\be \label{est}
T_c \approx 190 \ {\rm Mev}
\ee
(A more accurate
treatment which takes into account non-zero $m_{u,d}$ and also
the presence of other mesons in the heat bath gives practically
the same estimate as these two effects push $T_c$ in opposite
directions and practically cancel each other.)

\section{Properties of phase transition. The real World.}
\setcounter{equation}0

Making the estimate (\ref{est}), we tacitly assumed that the
phase transition is of the second order: only the derivative
$\partial<\bar q q>_T/\partial T$ but not $<\bar q q>_T$ is
discontinuous at the phase transition point. Let us discuss the
question whether this assumption is valid and under what
conditions.

On the theoretical side, the situation is similar to that in
pure YM case: not rigorous but suggestive arguments exist
indicating that the phase transition is of the second order for
2 massless flavors. When $N_f \geq 3$, the phase transition is
probably of the first order. The arguments are the following
\cite{PW}:

The starting point is the observation that, in theories
involving {\it scalar} fields,  phase transition of the first
order often occurs when the potential involves a cubic in fields
term. One can recall in the first place a cubic Van-der-Vaals
curve $P(\rho, T)$ which describe the first order water
$\leftrightarrow$ vapor phase transition. The simplest field
theory example is the theory of real scalar field with the
potential
\be \label{Vphi}
V(\phi) = \lambda(\phi^2 - v^2)^2 -
\mu \phi^3
\ee
Assume for simplicity $\mu \ll \lambda v$. At
$T=0$, the potential has one global minimum at $\phi \approx v +
3\mu/8\lambda$. At non-zero temperature, the term $\sim \lambda
T^2 \phi^2$ is added to the effective potential. At high
temperature $T \gg v$, the minimum occurs at $\phi = 0$. One can
be easily convinced that a {\it local } minimum at $\phi = 0$
appears at some temperature $T^*$ when the local minimum at
positive $\phi$ still exists. The latter disappears at some
larger temperature $T^{**}$. In a certain temperature range, two
minima of the potential, the old and the new one, coexist, one
being a metastable state with respect to the other.  This is
exactly the physical situation of the first order phase
transition.

Let us go back to $QCD$. A direct application of this reasoning
is not possible because the $QCD$ lagrangian does not involve
scalar fields. The effective chiral lagrangian (\ref{Lchi}) is
also of no immediate use because higher-derivative terms which
stand for dots cannot be neglected in the region close to
critical temperature. Suppose, however, that in the region $T
\sim T_c$ some other effective lagrangian in Ginzburg-Landau
spirit can be written which depends on the composite colorless
fields
\be \label{Fia}
\Phi_{ff'} = \bar q_{Rf} q_{Lf'}
  \ee
A general form of the effective potential which is invariant
under $SU_L(N_f) \otimes SU_R(N_f)$ is
\be
 \label{V3}
V[\Phi] \sim  g_1 {\rm Tr}\{ \Phi \Phi^\dagger\}
+ g_2 ({\rm Tr}\{ \Phi \Phi^\dagger\})^2 \nonumber \\
+ g_3{\rm Tr}\{ \Phi \Phi^\dagger \Phi \Phi^\dagger\}
+ g_4 (\det \Phi + \det \Phi^\dagger ) + \ldots
 \ee
(the coefficients may be smooth functions of $T$).  Now look at
the determinant term. For $N_f = 2$, it is quadratic in fields
while, for $N_f = 3$, it is cubic in fields and the effective
potential acquires the structure similar to Eq.(\ref{Vphi})
which is characteristic for the systems with first order phase
transition. A more refined analysis \cite{PW} shows that the first order
phase transition is allowed also for $N_f \geq 4$, but not for
$N_f =2$ where the phase transition is of the second order.

It is even possible to argue that, for $N_f \geq 3$, one has not one but
{\it two} phase transitions. The argument is based on the exact relation
for the spectral density of Euclidean Dirac operator $\rho(\lambda)$
in zero-temperature $QCD$ with $N_f$ massless flavours at small but nonzero
$\lambda$. It is possible to show that it involves a non-analytic term in
$\lambda$ at $N_f \geq 3$ \cite{SS}:
  \be
  \label{Stern}
\rho(\lambda) = \frac \Sigma \pi + \frac {\Sigma^2 (N_f^2 -4)}{32\pi^2 N_f
F_\pi^4} |\lambda| + o(\lambda^2)
  \ee
where $\Sigma = |<\bar q q>_0|$. At nonzero temperature, both $\rho(0)$ (i.e.
the chiral condensate) and the coefficient of $|\lambda|$ in $\rho(\lambda)$
are changed. Eventually (at $T \gg \mu_{hadr}$) they should vanish.
One can imagine a situation when the chiral condensate turns to
zero before the slope does. Or other way round.

If lattice people
would eventually observe two temperature phase transitions in
$QCD$ with 3 massless quark flavors, I would be happy, of
course. However, a canonical viewpoint that there is only one
transition but of the first order may also be true. Existing
lattice measurements favour this possibility.

Up to now, we discussed only pure YM theory and $QCD$ with
massless quarks. But the quarks have non-zero masses: $m_u
\approx$ 4 Mev, $m_d \approx$ 7 Mev, and $m_s \approx$  150
Mev \cite{GLmass}. The
question arises whether the non-zero masses affect the
conclusion on the existence or non-existence and the properties
of the phase transition.

The experimental (i.e. lattice) answer to this question appears
to be positive \cite{Columb}. In Fig. 2, a phase diagram of
$QCD$ with different values of quark masses $m_s$ and $m_u =
m_d$ is plotted.

\vspace{13cm}
Let us discuss different regions on this plot. When the quark
masses are large, quarks effectively decouple and we have pure
YM theory with $SU(3)$ gauge group where the phase transition is
of the first order. When all the quark masses are zero, the
phase transition is also of the first order. When masses are
shifted from zero a little bit, we still have a first order
phase transition because a finite discontinuity in energy and
other thermodynamic quantities cannot disappear at once when external
parameters (the quark masses) are smoothly changed.

But when all the masses are non-zero and neither are too small
nor too large, phase transition is absent. Notice the bold
vertical line on the left. When $m_u = m_d = 0$ and $m_s$ is not
too small, we have effectively the theory with two massless
quarks and the phase transition if of the second order. The
experimental values of quark masses (the dashed circle in Fig.
2) lie close to this line of second order phase transitions but
in the region where no phase transition occurs. It is the
experimental fact as measured in Ref. \cite{Columb}.

This statement conforms nicely with a semi-phenomenological
theoretical argument of ref. \cite{KK} which displays that even
{\it if} the first order phase transition occurs in QCD, it is
rather weak. The argument is based on a generalized
Clausius-Clapeyron relation. In college physics, it is the
relation connecting the discontinuity in free energy at the
first-order phase transition point with the sensitivity of the
critical temperature to pressure. The Clausius-Clapeyron
relation in $QCD$ reads
\be \label{KK}
  {\rm disc} <\bar q q>_T = \frac 1{T_c} \ \frac {\partial T_c}
{\partial m_q} {\rm disc} \ \epsilon
  \ee
where ${\rm disc}\ \epsilon$ is the latent heat.
The derivative $\frac {\partial T_c}{\partial m_q}$ can the
estimated from theoretical and experimental information of how
other essential properties of $QCD$ depend on $m_q$ and from the
calculation of $T$ - dependence of condensate at low temperature
in the framework of chiral perturbation theory (see Fig.1 and the
discussion thereof).  The
dependence on quark masses is not too weak. From that, assuming
 that the discontinuity in quark
condensate is as large as $<\bar q q>_0$ (which is not true, of course),
we get an estimate
$${\rm disc}\ \epsilon \ < 0.4 \ {\rm GeV/fm}^3$$
which is rather small compared to the characteristic free energy
density of the  system in the vicinity of $T_c \sim 190 {\rm MeV}$.

Thus, latent heat of the first order phase transition (assuming it is there)
must be small which means that the phase transition is likely to disappear
under a relatively small perturbation due to nonzero $m_s$.

The question is not yet completely resolved, and independent
lattice measurements are highly desirable.  Most probable is,
however, that, when temperature is changed, hadron gas goes over
to quark-gluon plasma and other way round without any phase
transition. There is, however, a sharp crossover in a narrow
temperature range which is similar in properties to a
second-order phase transition (the ``phase crossover'' if you
will).

  \section{Instantons and percolation.}
\setcounter{equation}0
In the analysis in previous two sections, we relied on the fact
that chiral symmetry {\it is} broken at zero temperature. It is
an experimental fact in real $QCD$, but it is important to
understand from pure theoretical premises {\it why} it is broken
and what is the mechanism of its restoration at higher
temperatures.

A completely satisfactory answer  to this question has not yet
been obtained. The problem is that $QCD$ at zero temperature is
a theory with strong coupling and it is very difficult (may be
impossible) to study the structure of $QCD$ vacuum state
analytically.  However, a rather appealing qualitative physical
picture exists which is based on the model of
instanton-antiinstanton liquid and on the analogy with the so
called percolation phase transition in doped superconductors
\cite{Shur}. We refer the reader to the Shuryak's book for the
detailed discussion and elucidate here only crucial points of
the reasoning.

The starting point is the famous Banks and Casher relation
\cite{Banks} connecting quark condensate to the mean spectral
density of Euclidean Dirac operator $\rho(\lambda)$ at $\lambda
\sim 0$. Let us explain how it is derived. Consider the
Euclidean fermion Green's function $<q(x) \bar q(y)>$ in a
particular gauge field background. Introduce a finite Euclidean
volume $V$ to regularize theory in  the infrared.  Then the
spectrum of massless Dirac operator is discrete and enjoys the
chiral symmetry: for any eigenfunction $\psi_n(x)$ satisfying
the equation $\not\!\!{\cal D} \psi_n = \lambda_n \psi_n$ , the
function $\tilde \psi_n = \gamma^5 \psi_n$ is also an
eigenfunction with the eigenvalue $\tilde \lambda_n = -
\lambda_n$.

The idea is to use the spectral decomposition of the fermion
Green's function with a small but non-zero quark mass
\be
\label{Green} <q(x) \bar q(y)> \ = \ \sum_n \frac {\psi_n(x)
\psi_n^\dagger (y)}{i\lambda_n - m}
  \ee
Set $x=y$ and integrate over $d^4x$. We have
\be \label{sum}
V<\bar q q> = - m \sum_{\lambda_n > 0} \frac 1{\lambda_n^2 +
m^2}
\ee
 where the chiral symmetry of the spectrum has been used
and the contribution of the zero modes $\lambda_n = 0$ has been
neglected (it is justified when the volume $V$ is large enough
\cite{LS}).  Perform the averaging over gauge fields and take
{\it first} the limit $V \to \infty$ and {\it then} the limit $m
\to 0$. The sum can be traded for the integral:
\be
\label{Banks} <\bar q q> = - m \int \frac
{\rho(\lambda)}{\lambda^2 + m^2} d\lambda = - \frac 1\pi \rho(0)
\ee
 The rightmost-hand-side of Eq.(\ref{Banks}) is only the
non-perturbative $m$-independent part of the condensate  .
There is also a perturbative ultraviolet-divergent piece
$\propto m\Lambda_{ultr}^2$ which is proportional to the quark
mass, is related to large eigenvalues $\lambda$ and is of no
concern for us here.

Thus, the non-perturbative part of the quark condensate which is
the order parameter of the symmetry breaking is related to small
eigenvalues of Euclidean Dirac operator.  There should be a lot
of them --- a characteristic spacing between levels is $\delta
\lambda \sim 1/(|<\bar q q>|V)$ which is much less than the
characteristic spacing $\delta \lambda \sim 1/L$ for free
fermions.

The question is what is the physical reason for these small
eigenvalues to appear. As far as we know, the first pioneer
paper where a mechanism for generating small eigenvalues was
proposed is Ref.\cite{Flor} where small eigenvalues appeared as
zero modes of monopole-like gauge field configurations. The
disadvantage of this model is that the monopole configurations
are static whereas it is natural to expect that characteristic
gauge fields contributing to the Euclidean path integral at
$T=0$ are more or less symmetric in all four directions with no
particular axis being singled out. The model of
instanton-antiinstanton liquid formulated in \cite{Diak} (see in particular
Ref. [28b] where the mechanism for spontaneous chiral symmetry breaking
was suggested) and
developed later in \cite{Shur1} is much better in this respect.

The basic assumption of the model is that a characteristic gauge
field contributing in  $QCD$ path integral is a medium of
instantons and instantons as shown in Fig.3.  It is not a ``gas''
of Callan, Dashen, and Gross \cite{CDG} because the interaction
between instantons and antiinstantons bringing about a
short-range correlations between instanton positions and
orientations cannot be neglected. A ``liquid'' is a more proper
term.

\vspace{1cm}

\begin{picture}(90,55)
\put(10,10){\circle{15}}
\put(9.5,8){{\Large I}}
  \put(30,20){\circle{15}}
  \put(28.5,18){{\Large A}}
\put(29,36){\circle{15}}
\put(28.5,34){{\Large I}}
  \put(12,30){\circle{15}}
  \put(10.5,28){{\Large A}}
\put(48,34){\circle{15}}
\put(47.5,32){{\Large I}}
\put(45,7){\circle{15}}
\put(44.5,5){{\Large I}}
  \put(60,10){\circle{15}}
  \put(58.5,8){{\Large A}}
  \put(63,30){\circle{15}}
  \put(61.5,28){{\Large A}}
\put(82,20){\circle{15}}
\put(81.5,18){{\Large I}}
\put(58,48){\circle{15}}
\put(57.5,46){{\Large I}}
  \put(44,50){\circle{15}}
  \put(42.5,48){{\Large A}}
\put(12,45){\circle{15}}
\put(11.5,43){{\Large I}}

\end{picture}
\vspace{0.6cm}

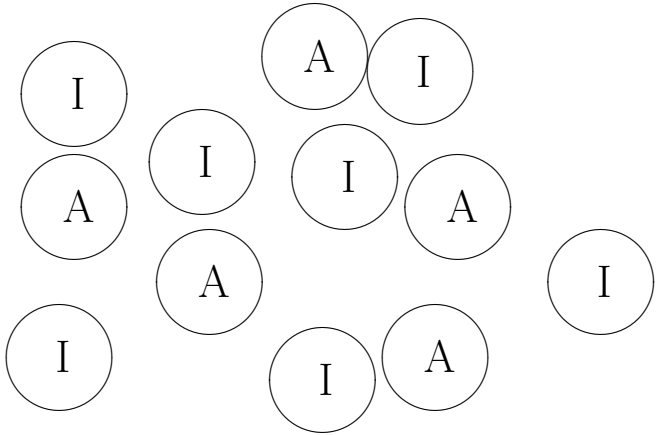
{\bf Fig. 3}. Instanton-antiinstanton liquid.

\vspace{0.6cm}

Each individual instanton and antiinstanton involves a fermion
zero mode \cite{Hooft}. Assuming the constant density of
quasi-particles $\propto \mu_{hadr}^4$, the total number of zero
modes in the Euclidean volume $V$ is $N \sim V\mu_{hadr}^4$.
However, these are not {\it exact} zero modes. They are shifted
from zero due to interaction between instantons and
antiinstantons (a nonzero overlap between individual instanton
and antiinstanton zero modes. Assuming their uniform spreading in the range of
eigenvalues $\Delta \lambda \sim \mu_{hadr}$, the volume density
of {\it quasi-zero modes} is $\rho(0)
\sim N/(V\Delta \lambda) \sim \mu_{hadr}^3$. Due to
Eq.(\ref{Banks}), a non-zero quark condensate appears
\footnote{The assumption of quasi-uniform spreading of eigenvalues is not so
innocent. It probably holds only in the theory with several light dynamical
quarks, but not in the quenched theory ($N_f = 0$) where it is natural to
expect a singular behaviour of the spectral density near zero: $\rho(\lambda)
\sim 1/\lambda$ so that the ``fermion condensate'' (i.e. the vacuum
expectation value $<\bar q q>_0$ where quark fields are treated as external
sources) is infinite \cite{SMvac}.}.

This picture is rather similar to what happens in a doped
superconductor with high enough doping. When a characteristic
distance between individual atoms of the admixture is not large,
the wave functions of outer electrons of these atoms overlap,
and the electrons can jump from site to site. If the set of
atoms of admixture with a noticeable overlap of wave functions
forms a connected network in the space, the electrons can travel
through this network at large distances and the specimen is a
{\it conductor}.  Note that it is not a standard metal mechanism
of conductivity when the medium is a crystal, has the long-range
order, and the electron wave functions are periodic Bloch waves.
Here the distribution of the dope whose electrons are
responsible for conductivity is stochastic and wave functions
are complicated.  The essential is that they are {\it
delocalized}.

Thus, one can say that the vacuum of $QCD$ is the ``conductor'' in
a certain sense. For sure, there is no conductivity of anything
in usual Minkowski space-time. Only the Euclidean vacuum
functional has ``conducting'' properties. In principle, one can
introduce formally the fifth time and write an analog of Kubo
formula for conductivity in $QCD$, but the physical meaning of
this ``conductivity'' is not clear.  It is sufficient to say that,
in a characteristic Euclidean gauge field background, the
eigenfunctions of Dirac operator corresponding to small
eigenvalues are delocalized.

What happens if we heat the system ? The effective coupling
constant $g^2(T)$ decreases, the action of individual instantons
$S = 8\pi^2/g^2(T)$ increases, and the density of
quasi-particles $\propto \exp\{-S\}$ decreases.

Let us look first at the doped superconductor when we decrease
the density of admixture. Below some critical density, the set
of atoms with essential overlapping of wave functions does not
form a connected network in 3-dimensional space anymore.  The
electrons can no longer travel far through this network, wave functions become
localized, and the
specimen is an insulator. This is called the percolation phase
transition (see e.g.
\cite{Shkl} for detailed discussion).

Likewise, there is a critical temperature in $QCD$ above which
instantons and antiinstantons do not form anymore a connected
cluster with an essential overlap of individual fermion zero
modes [what overlap is ``essential'' and what is not  is a
numerical question.  For condensed matter systems (but not for
$QCD$ in this context) a computer estimates for the critical
admixture density has been performed]. At high temperatures, few
remaining quasi-particles tend to form ``instanton-antiinstanton
molecules'' (See Fig.4). The individual zero modes are not {\it
spread out} uniformly in the range $\Delta \lambda \sim
\mu_{hadr}$ as is the case at zero temperature where instantons
and antiinstantons form an infinite cluster, but are just {\it
shifted} by the value $\sim \mu_{hadr}$ due to interaction in
individual molecules. Small eigenvalues in the spectrum of Dirac
operator are absent and the fermion condensate is zero
\cite{Shur}.

\begin{picture}(90,85)
\put(10,10){\circle{15}}
\put(9.5,8){{\Large I}}
  \put(27,11){\circle{15}}
  \put(25.5,9){{\Large A}}
\put(50,58){\circle{15}}
\put(49.5,56){{\Large I}}
  \put(52,42){\circle{15}}
  \put(50.5,40){{\Large A}}
\put(85,8){\circle{15}}
\put(84.5,6){{\Large I}}
  \put(69,11){\circle{15}}
  \put(67.5,9){{\Large A}}
\put(130,65){\circle{15}}
\put(129.5,63){{\Large I}}
  \put(146.5,64){\circle{15}}
  \put(145,62){{\Large A}}

\end{picture}

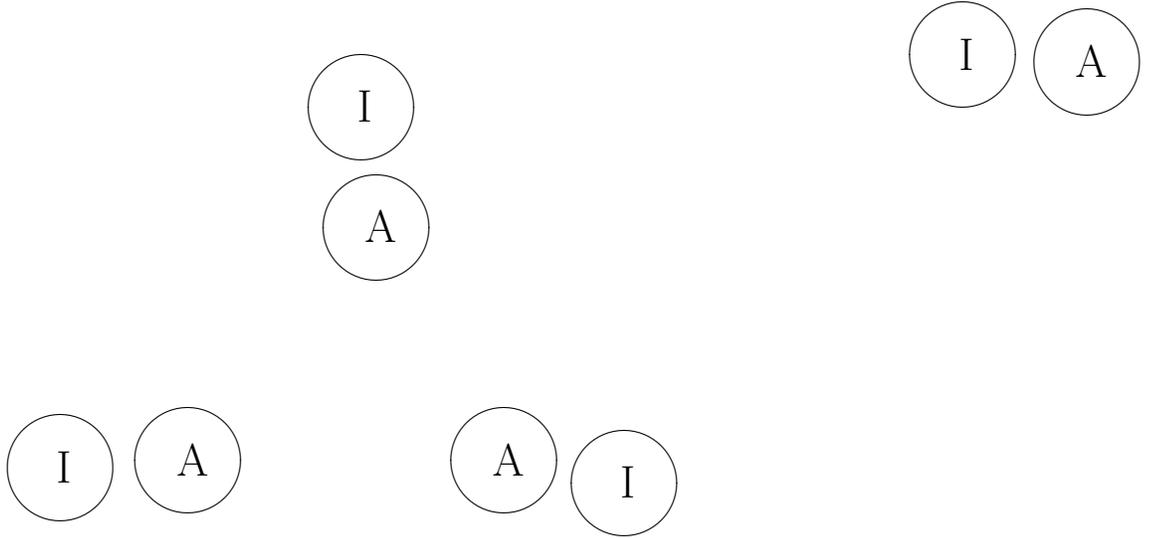
{\bf Fig. 4}. Gas of instanton-antiinstanton molecules (high $T$).
\vspace{0.6cm}

Of course, this picture is too heuristic and qualitative.  A
serious quantitative study of fermion eigenvalues and
eigenfunctions at non-zero temperature, and especially in the
region $T \sim T_c$ has not yet been done. It is the task (a
very interesting and important one) for future explorers.

It is worthwhile to emphasize once more that this scenario of percolation
phase transition leading to the molecular high-temperature phase is expected
to hold only at $N_f \geq 2$. For $N_f =1$ with arbitrary small but nonzero
fermion mass, molecules get ionized and the ``medium'' presents a very dilute
instanton-antiinstanton gas --- the instanton density involves a product of two
small factors: $\exp\{-8\pi^2/g^2(T)\}$ and the fermion mass $m$.
Differentiating ${\rm log} Z$ over $m$ and sending $m$ to zero, one gets a
small but non-zero quark
condensate \cite{GPJ,KY}. Cf. the analogous situation in the Schwinger model
\cite{Schwinst}

\section{Disoriented chiral condensate.}
\setcounter{equation}0

When we talked in previous sections about ``experimental'' tests
of theoretical predictions, we meant  numerical lattice
experiment. It is the unfortunate reality of our time that the
feedback between  theory and real laboratory experiment has
drastically deteriorated: what is interesting from theoretical
viewpoint cannot very often be measured in laboratory and what
can be measured is not interesting.

However, speaking of the particular problem of the phase
transition in $QCD$ associated with chiral symmetry restoration,
an intriguing possibility exists that a direct experimental
evidence for such a transition can be obtained at the
high-energy heavy ion collider RHIC which is now under
construction. After a head-on collision of two energetic heavy
nuclei, a  high temperature hadron ``soup'' is created.

We do not call this soup the quark-gluon plasma because, even at
RHIC energies, the temperature would not be high enough to
provide a sufficient smallness of the effective coupling
$g^2(T)$ and to make the perturbation theory over this parameter
meaningful. What is important, however, is that, at RHIC
energies, the temperature of the soup would be well above the
estimate (\ref{est}) for the phase transition temperature. The
high-temperature state created in heavy nuclei collision would
exist for a very short time after which it expands, is cooled
down and decays eventually into  mesons.

Let us look in more details at the cooling stage. At high
temperature, the fermion condensate is zero. Below phase
transition, it is formed and breaks spontaneously chiral
symmetry. This breaking means that the vacuum state is not
invariant under the chiral transformations (\ref{chi}) and a
direction in isotopic space is distinguished. What particular
direction --- is a matter of chance. This direction is specified
by the condensate matrix
\be \label{condmat}
\Sigma_{f f'} = <\bar q_{Lf} q_{Rf'}>
  \ee
For simplicity, we have assumed up to now that
\be \label{diag}
\Sigma_{ff'} = -\Sigma \delta_{ff'}
  \ee
, but any unitary matrix can be substituted for $\delta_{ff'}$
(of course, it can be brought back in the form $\delta_{ff'}$ by
a chiral transformation ).  In different regions of space,
cooling occurs independently and directions of condensate are
not correlated. As a result, domains with different directions
of condensate shown in Fig.5 are formed (cf. cooling down of a
ferromagnetic below the Curie point).

In our World, we do not observe any domains, however. The
direction of the condensate in all spatial points is identical.
This is a consequence of the fact that $u$- and $d$- quarks have
non-zero masses which break chiral symmetry explicitly, the
vacuum energy involves a term
\be \label{EMSig} E_{vac} \sim
{\rm Tr} \{{\cal M}^\dagger \Sigma \} + {\rm c.c.}
\ee and the
only true vacuum state is (\ref{diag}) (in the basis where the
quark mass matrix ${\cal M}$ is diagonal).

However, the masses of $u$- and $d$- quarks are rather small and
one can expect that the domains with ``wrong'' direction of the
condensate are sufficiently developed during the cooling stage
before they eventually decay into true vacuum (\ref{diag}) with
emission of pions.
\footnote{Fig.5 implies the existence of several domains
and describes better the physical situation immediately after
the ``phase crossover'' in early Universe. Probably, the size of
the hot fireball produced in collision of two nuclei is too
small and the cooling occurs too fast for several domains to be
developed. The popular ``baked Alaska'' scenario \cite{Bj} implies
the formation of only one domain with (generally) wrong flavour
orientation. }

This is a crucial assumption. A theoretic estimate of the
characteristic size of domains they reach before decaying is
very difficult and there is no unique opinion on this question
in the literature. But if this assumption is true, we can expect
to observe a very beautiful effect \cite{Bj}.  From the true
vacuum viewpoint, a domain with disoriented $\Sigma_{ff'}$ is a
classical object --- kind of a ``soliton'' (parentheses are put
because it is not stable) presenting a {\it coherent}
superposition of many pions. The mass of this quasi-soliton is
much larger than the pion mass. The existence of such multipion
coherent states was discussed long ago in pioneer papers
\cite{Ans} but not in relation with thermal phase transition.

Eventually, these objects decay into pions. Some of the latter
are neutral and some are charged. As all isotopic orientations
of the condensate in the domains are equally probable, the {\it
average} fractions of $\pi^0$, $\pi^+$, and $\pi^-$ are equal:
$<f_{\pi^0}> = <f_{\pi^\pm}> =
\frac 13$ as is also the case for incoherent production of
pions in, say, $pp$ collisions where no thermalized high-$T$
hadron soup is created.

But the {\it distribution} $P(f)$ over the fraction of, say,
neutral pions is quite different in the case of incoherent and
coherent production. In incoherent case, $P(f)$ is a very narrow
Poissonic distribution with the central value $<f_{\pi^0}> =
1/3$. The events with $f_{\pi^0} = 0$ or with $f_{\pi^0} = 1$
are highly unprobable: $P(0) \sim P(1) \sim
\exp\{ - C N \}$ where $N \gg 1$ is the total number
of pions produced.

For coherent production, the picture is quite different.
$\Sigma_{ff'}$ is proportional to a $SU(2)$ matrix.  Factorizing
over $U(1)$, one can define a unit vector in isotopic space $\in
S^2$. The fraction of $\pi^0$ produced would be just $f = \cos
^2 \theta$ where $\theta$ is a polar angle on $S^2$.  The
probability to have a particular polar angle $\theta$ normalized
in the interval $0 \leq \theta
\leq \pi/2$ [ the angles $\theta > \pi/2$ do not bring about
anything new as $f(\pi - \theta) = f(\theta)$ ] is $P(\theta) =
\sin \theta$.  After an elementary transformation, we get a
normalized probability in terms of $f$:
\be \label{Pf}
P(f) df =
\frac {df}{2 \sqrt{f}}
\ee As earlier, $<f> = 1/3$, but the
distribution in $f$ is now wide and the values $f = 0$ and $f =
1$ are quite probable.

Thus, a hope exists that in, experiments with heavy ion
collisions at RHIC, wild fluctuations in the fractions of
neutral and charged pions would be observed. That would be a
direct experimental indication that a quasi-phase-transition
occurs where domains of disoriented chiral condensate of
noticeable size are developed in a cooling stage. One can recall
in this respect mysterious Centauro events with anomalously
large fraction of neutral or of charged particles observed in
cosmic ray experiments \cite{Cent}. Who knows, may be that {\it
was} the first experimental observation of the $QCD$ phase
transition.

\section{Aknowlegdements}
It is a pleasure to thank the organizers of the Enrico Fermi School
for the kind hospitality. This work was partially supported
by the INTAS grant 93-0283.

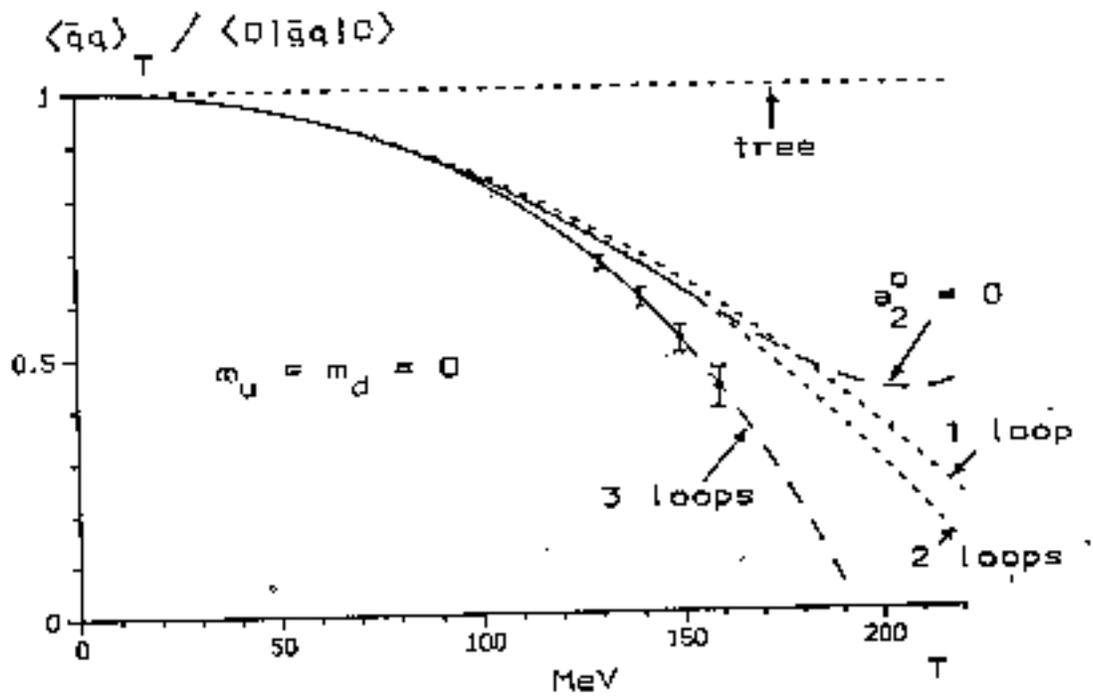
\end{document}